\documentclass[onecollarge,natbib]{svjour2}
\bibpunct{[}{]}{;}{n}{}{,} 
\smartqed  
\usepackage{graphicx}
\usepackage[utf8]{inputenc}
\usepackage{amsmath}      
\journalname{Few-Body Systems}
\begin{document}

\title{$N^*$ Resonances in Lattice QCD from (mostly) Low to (sometimes) High Virtualities
}

\titlerunning{N* resonances on lattice\dots}        

\author{David G.~Richards}


\institute{Jefferson Lab\\
12000 Jefferson Avenue\\  
Newport News, VA 23606\\
USA\\
              Tel.: +1-757-269-7736\\
              \email{dgr@jlab.org}           
}

\date{Received: date / Accepted: date}

\maketitle

\begin{abstract}
I present a survey of calculations of the excited $N^*$ spectrum in
lattice QCD.  I then describe recent advances aimed at extracting the
momentum-dependent phase shifts from lattice calculations, notably in
the meson sector, and the potential for their application to baryons.  
I conclude with a discussion of calculations of the electromagnetic
transition form factors to excited nucleons, including calculations at
high $Q^2$.
\end{abstract}

\section{Introduction}
\label{intro}
Lattice gauge theory provides a means of \textit{solving} QCD in the
low-energy regime, and thereby has a key role in hadronic and nuclear
physics. The calculation of the low-lying spectrum, the masses of
hadrons stable under the strong interactions, has long been a
benchmark calculation for lattice QCD since it provides a
confrontation between lattice QCD and precisely determined
quantities in experiment.  Such calculations are demanding, since they
require a high degree of control over the systematic uncertainties
inherent to lattice calculations, namely those arising from the finite
volume in which they are performed, non-zero lattice spacing, and,
finally, the need until recently to extrapolate from unphysical $u$
and $d$ quark masses to the physical light quark masses.

The focus of this talk is the spectrum of excited-state nucleons, and
on the determination of their properties and photoproduction
mechanisms.  In fact, my emphasis will very much be on the former
since it is a prerequisite for the latter; we will hear elsewhere in
this workshop of recent theoretical developments aimed at a rigorous
determination of hadronic matrix elements that involve multi-hadron
states\cite{raul}.  I will begin by reminding us briefly of the
formalism of lattice QCD, and note some of the computational
challenges.  I will point out some recent successes at looking at the
properties of stable hadrons such as the nucleon, before proceeding to
describe how the spectrum of excited states can be determined on the
lattice.  In particular, I will focus on the use of the variational
method, and how that can provide important insights both into the
masses of their states, and their quark and gluon structure. I will
then describe how resonances are treated in a lattice calculation, and
the determination of the momentum-dependent phase shifts.  I will
conclude with a discussion of lattice calculations of transition form
factors, and a summary.

\section{Lattice QCD and the Properties of Ground-state Hadrons}
Lattice QCD is Quantum Chromodynamics formulated on a Euclidean
space-time lattice, with the ``lattice spacing'' fulfilling the
r\^{o}le of a hard, ultraviolet cut-off.  An ``observable'' in lattice QCD
can be calculated through the evaluation of the discretized path
integral
\begin{equation}
  \langle {\cal O} \rangle = \frac{1}{\cal Z} \prod_{x,\mu}
  dU_{\mu}(x) \prod_{x} d\psi(x) \prod_{x} d\bar{\psi}(x) {\cal O}(U,
    \psi, \bar{\psi}) e^{-S(U, \psi, \bar{\psi})},
\end{equation}
where $U_{\mu}(x)$ are $SU(3)$ matrices representing the gluonic
degrees of freedom, and $\psi, \bar{\psi}$ are Grassmann variables
representing the quarks.  Writing the QCD action in terms of the pure
gauge and fermion components, we have
\[
S(U,\psi,\bar{\psi}) = S_g(U) + \bar{\psi} M(U) \psi,
\]
whence we can integrate the fermion degrees of freedom to yield
\begin{equation}
  \langle{\cal O} \rangle = \frac{1}{\cal Z} \prod_{x,\mu} dU_{\mu}(x)
 {\cal O}(U) \det M(U) e^{- S_g(U)}.\label{eq:path}
\end{equation}
The need to formulate the theory in Euclidean space is now clear:
eqn~\ref{eq:path} can be estimated using the \textit{importance
  sampling} methods familiar to statistical physics, since the action
$S_g(U)$, and indeed the determinant, are real. A lattice gauge
calculation therefore consists, firstly, of generating an ensemble of
equilibrated gauge configuration
$\{U^n_{\mu}(x): n = 1,\dots,N_{\rm cfg}\}$ distributed according to
\begin{equation}
P(U) \propto \det M(U) e^{-S_g(U)} \label{eq:distrib}
\end{equation}
and then calculating the expectation value of the operator ${\cal O}$ on
\begin{equation}
\langle {\cal O} \rangle = \frac{1}{N_{\rm cfg}} \sum_{n=1}^{N_{\rm
      cfg}} {\cal O}(U^n, G[U^n])\label{eq:observ}
\end{equation}
where $G[U^n]$ represent quark propagators computed on the gauge
field $U^n_{\mu}(x)$.  The generation of the gauge fields in
eqn~\ref{eq:distrib} is a \textit{capability} computing task in which
the whole, or at least a sizable proportion, of a leadership-class
computer has to be devoted to a single sequence of Monte Carlo calculations to
yield a thermalized distribution. The evaluation of $\det M(U)$
is dominant, since any local change to gauge linke $U_{\mu}(x)$
requires the evaluation of the determinant over the whole lattice, in
contrast to a pure-gauge calculation where the change in $S_g(U)$ can
be evaluated locally.  The evaluation of
eqn~\ref{eq:observ} is inherently a \textit{capacity} computing task,
since the calculation of ${\cal O}(U^n)$ can be performed on each
configuration independently.  For many of the calculations
described below, the integrated cost of these computations can far
exceed those of eqn~\ref{eq:distrib}, but these can exploit
cost-optimized clusters and GPU-accelerated clusters.

The spectrum of QCD is determined
through the exponential decay of two-point correlation functions
\begin{equation}
  C(t) = \sum_{\vec{x}} \langle 0 \mid {\cal O}(\vec{x},t) \bar{\cal O}(0) \mid 0
  \rangle \longrightarrow \sum_n A_n e^{-E_n t}, \label{eq:2pt}
\end{equation}
where $\cal O$ is an interpolating operator, say for the nucleon, and the
$E_n$ and $A_n$ are the energies and residues of all states $n$ with
$\langle n \mid \bar{\cal O} \mid 0 \rangle \ne 0$; the energies $E_n$ are
real, reflecting the formulation in Euclidean space.  The masses of
the low-lying states of the spectrum, that is states stable under the
strong interaction, are determined from the leading exponential in
eqn~\ref{eq:2pt}.

As noted earlier, calculations that can directly confront experiment
require control over a variety of systematic uncertainties.  A
prominent example of such a calculation, for states composed of the
$u,d$ and $s$ quarks, was that of the BMW
collaboration\cite{Durr:2008zz}, and they have recently extended their
work through the inclusion of QED computations to provide a
calculation of the proton-neutron mass difference, as well as that for
other hadrons\cite{Borsanyi:2014jba}.  Beyond the low-lying spectrum,
calculations are also being performed of the matrix elements of the
nucleon, including those for the (space-like) electric and magnetic
form factors, and for the moments of parton distributions and
Generalized Parton Distributions, initially for the isovector
properties, but now extended to include the effects of the sea quarks;
for a recent review,
see ref.~\cite{Constantinou:2015agp}. For each of these quantities,
calculations in Euclidean space can be directly related to quantities
in Minkowski space.  None-the-less, even for ground-state hadrons,
there are limitations to properties that can be straightforwardly
calculated, notably the matrix elements of the quark bilinears
separated along the light cone that give rise to the (generalized)
parton distribution functions and transverse-momentum-dependent
distributions, and methods have been
proposed\cite{Ji:2013dva,Xiong:2013bka,Ma:2014jla} and indeed applied,
to circumvent these issues in Euclidean-space calculations.

\section{The excited-state spectrum in lattice QCD}
The first challenge in the study of the excited-state spectrum is to
effectively determine the subleading exponentials in
eqn~\ref{eq:2pt}.  A robust way of doing so is
by means of the variational method, whereby we
compute a matrix of correlation functions
\begin{equation}
C_{ij}(t) = \sum_{\vec{x}} \langle 0 \mid {\cal O}^{J^P}_i(\vec{x},t)
\bar{\cal O}^{J^P}_j(\vec{0},0)
\mid 0 \rangle \longrightarrow \sum_n A^n_{ij} e^{- E_n t},\label{eq:matrix}
\end{equation}
where $\{ {\cal O}^{J^P}_i: i = 1,\dots,N\}$ is a basis of operators, each having
common quantum numbers.  We now solve the generalized eigenvalue equation
\begin{equation}
C(t) u(t,t_0) = \lambda(t,t_0) C(t_0) u(t,t_0)
\end{equation}
yielding a set of real eigenvalues $\{ \lambda_n(t,t_0): n =
1,\dots,N\}$ with corresponding eigenvectors $\{ u^n(t,t_0): n =
1,\dots,N\}$, where, for sufficiently large $t$, we have $\lambda_0
\ge \lambda_1 \ge \dots$.  These delineate the different states
\begin{equation}
\lambda_n(t,t_0) \rightarrow (1 - A) e^{-E_n (t - t_0)} + A e^{ - (E_n
  + \Delta E_n) (t - t_0)}.
\end{equation}

The application of the variational method replies on the construction of
a suitable basis of interpolating operators.  In the case of the
nucleon, there are only three local interpolating operators that can
be constructed from three quark fields, such that at most three
energies can be determined:
\begin{eqnarray}
  {\cal O}^{1/2} =
  \begin{cases} (u C \gamma_5 d) u \\
    (u C d) \gamma_5 u  \\
    (u C \gamma_4 \gamma_5 d) u \\
  \end{cases}. \label{eq:Ninterp}
\end{eqnarray}
The basis of operators can be extended to include quasi-local quark
operators through the use of gauge-invariant smearing whereby a quark
field $\psi(\vec{x},t) \longrightarrow \tilde{\psi}(\vec{x},t) =
\sum_{\vec{y}} L(\vec{x},\vec{y}) \psi(\vec{y})$ where
$L(\vec{x},\vec{y})$ is a gauge-covariant operator, such as the
inverse of a three-dimensional Laplacian; such a construction does not
alter the angular-momentum structure of a nucleon operator.  There
have been many studies in the last few years that have aimed at
extracting the excited-state nucleon spectrum of both parities through
the use of the variational
method\cite{Allton:1993wc,Burch:2004he,Burch:2006dg,Mahbub:2009nr,Mahbub:2010rm,Mahbub:2010me,Mahbub:2012ri},
as well as through techniques such as the sequential-Bayes method
aimed at delineating subleading terms from only a single
correlator\cite{Mathur:2003zf}.  A basis of operators with different
smearing radii for the quark fields can capture the radial structure
of the nucleons, and indeed allow for nodes in the wave function.
However, it appears incomplete in that it does not capture the orbital
structure. To do so, and indeed to
study states of spin higher than $3/2$, requires
operators that are non-local, either by displacing one or more of the
quarks, or equivalently, through the use of covariant
derivatives acting on the quark fields.  Such constructions, in which
the bases of operators were designed in the first instance to satisfy
the symmetries of the lattice, were introduced in
refs.~\cite{Basak:2005ir} and \cite{Basak:2005aq}, and applied in
refs.~\cite{Basak:2007kj,Bulava:2009jb,Bulava:2010yg}, providing, for
the first time, access to states of spin $5/2$ and higher.

The lack of rotational symmetry introduced through the discretisation
onto a finite space-time lattice has the consequence that angular
momentum is no longer a good quantum number at any finite spacing.  We
find in the meson sector that a remarkable degree of rotational symmetry in
operator overlaps is realized at the hadronic scale, enabling the
``single-particle'' spectrum to be classified according to the total
angular momentum of the states\cite{Dudek:2009qf,Dudek:2010wm}.  We
exploit this observation in constructing a basis of interpolating
operators for baryons.  We begin by expressing continuum baryon interpolating
operators of definite $J^P$ as\cite{Edwards:2011jj,Edwards:2012fx}
\begin{equation}
{\cal O}^{J^P} \sim  \left( F_{\Sigma_F} \otimes (S^{P_s})_{\Sigma_S}^n \otimes D^{[d]}_{L,
      \Sigma_D} \right)^{J^P},
\end{equation}
 where $F$, $S$ and $D$ are the flavor, Dirac spin and orbital angular
 momentum parts of the wave function, and the $\Sigma$'s express the
 corresponding permutation symmetry: Symmetric (S), mixed-symmetric
 (MS), mixed anti-symmetric (MA), and anti-symmetric (A).
 
\begin{figure*}
\centering
  \hspace*{-0.5cm}\includegraphics[width=0.58\textwidth]{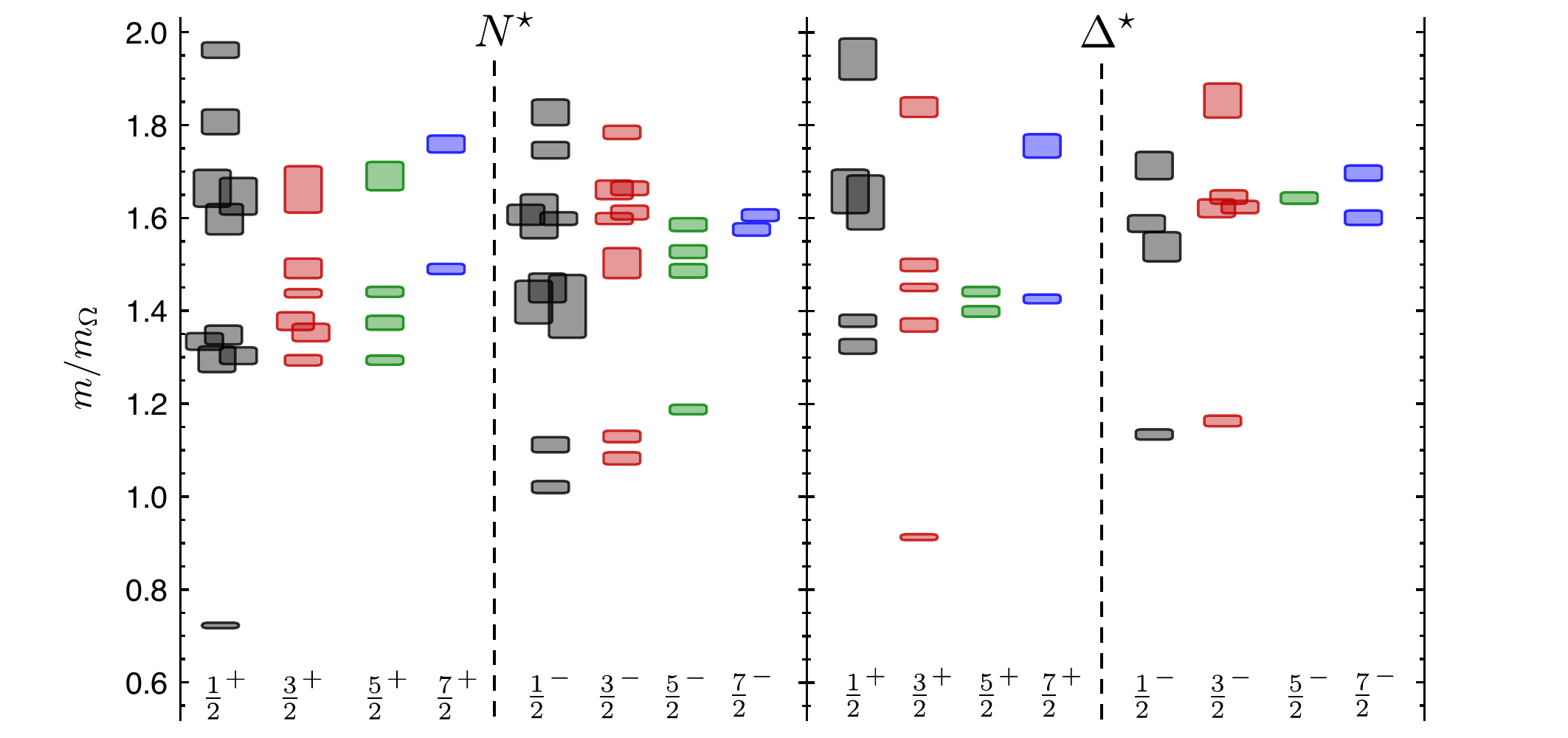}\includegraphics[width=0.48\textwidth]{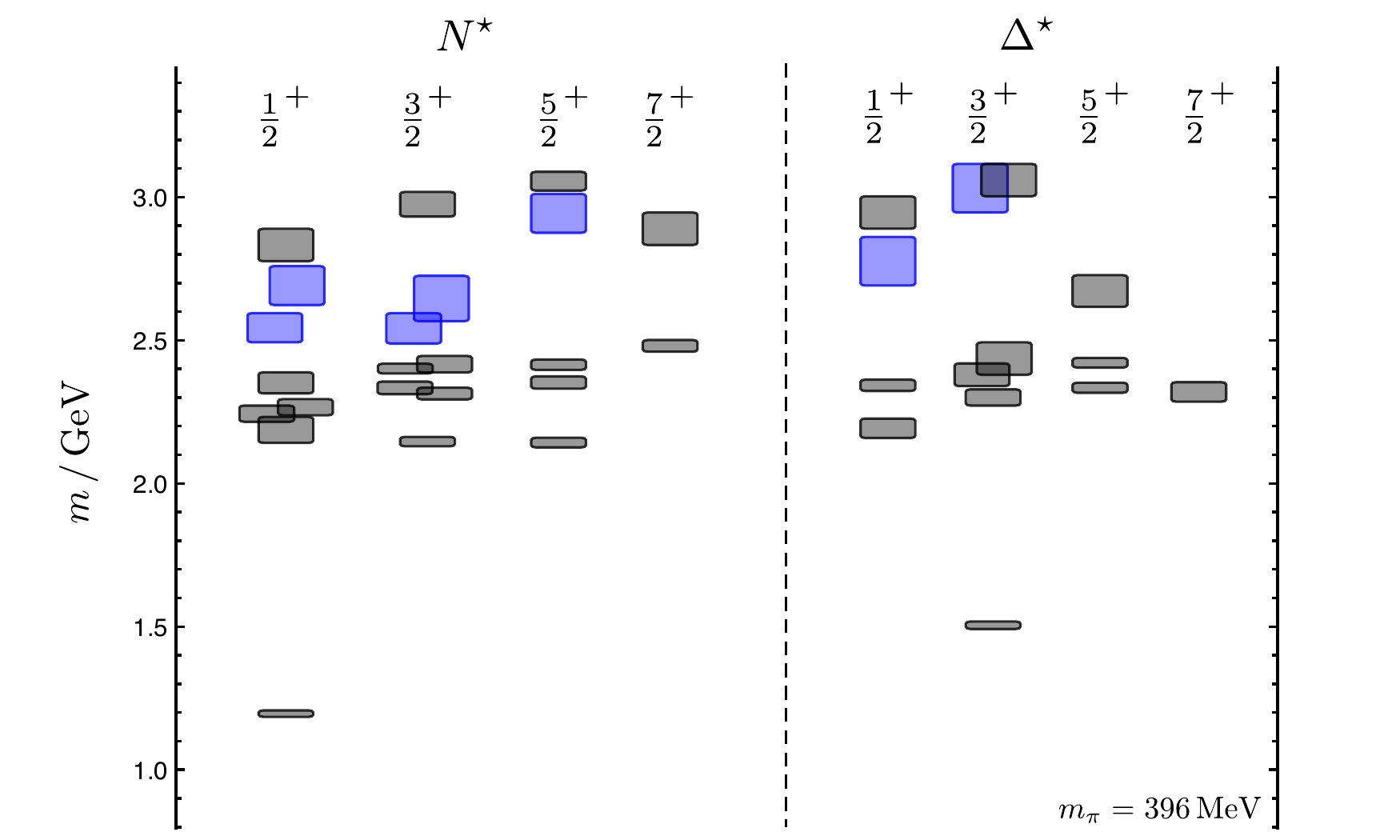}
\caption{The left-hand plot shows the spin-identified excited nucleon
  and $\Delta$ spectra in units of the $\Omega$ mass obtained on a
  $16^3\times 64$ anisotropic clover lattice with the strange quark
  mass at its physical value, and the light-quark masses corresponding
  to a pion mass of $396~{\rm MeV}$\protect\cite{Edwards:2011jj}.  The
  right-hand plot shows the positive-parity spectrum for both the
  nucleon and $\Delta$ in physical units using an operator basis with
  the ``hybrid'' operators included\cite{Dudek:2012ag}.  The
  additional states that couple predominantly to the hybrid-type
  operators are shown in blue.}
\label{fig:spin_840}       
\end{figure*}

 Non-zero orbital angular momentum is introduced through the use of
 gauge-covariant derivatives, written in a circular basis and acting
 on the quark fields.  In the notation above, $D^{[d]}_{L, \Sigma_D}$
 corresponds to an orbital wave function constructed from $d$
 derivatives, and projected onto orbital angular momentum $L$. In the
 calculations described here, up to two covariant derivatives are
 employed, enabling orbital angular momentum up to $L = 2$ to be
 accessed, and therefore states up to spin $7/2$ to be studied; the
 spin-identified spectrum obtained using such a basis of operators is
 shown as the left-hand plot in Figure~\ref{fig:spin_840}.  The
 spectrum reveals a counting consistent with that of a $qqq$
 non-relativistic quark model, and richer than certain quark-diquark pictures.

A faithful extraction of the spectrum is contingent on having a
sufficiently complete basis of operators. In the analysis described
above, one particular operator is absent, namely the mixed-symmetric
combination $D^{[2]}_{L = 1, M}$, the commutator of two covariant
derivatives projected to $L = 1$, that corresponds to a
chromo-magnetic field that would vanish for trivial gauge field
configuration; operators with this construction we identify as
\textit{hybrid} operators, associated with a manifest gluon
content\cite{Dudek:2012ag}.  The right-hand plot of
Figure~\ref{fig:spin_840} is the positive-parity nucleon and $\Delta$
spectrum on the same ensemble as that used in the left-hand plot,
showing the additional states appearing when this more complete basis
of operators is employed.  The conclusion is that, with the addition
of these operators, the counting of states is still richer than the
quark model, with the additional states attributable to a coupling to
a colour-octet gluonic excitation.

\begin{figure}
\centering
  \includegraphics[width=0.5\textwidth]{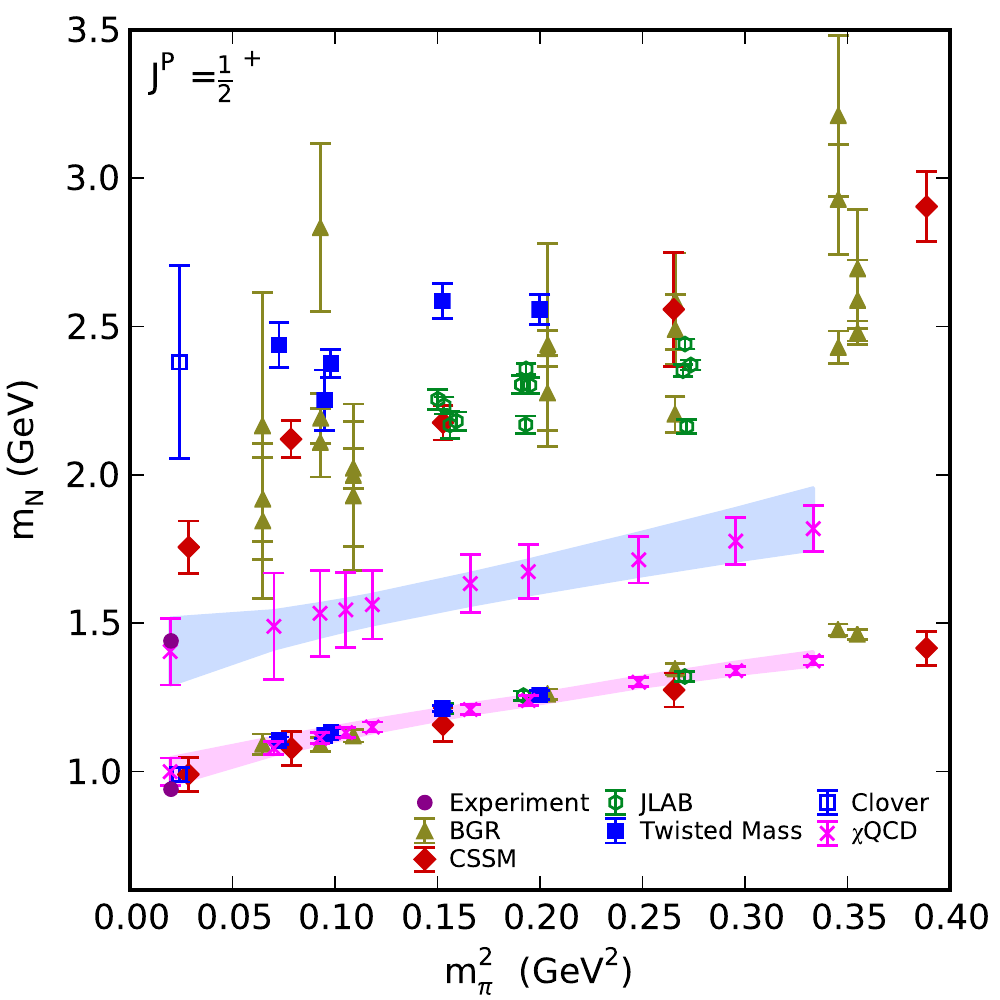}
\caption{Calculations of the lowest-lying positive parity excitation
  of the nucleon, taken from reference~\protect\cite{Liu:2014jua}. The
  $\chi$QCD results are obtained using overlap fermions on an ensemble
  generated using domain-wall fermions.}
\label{fig:roper}       
\end{figure}

One feature of the experimentally observed nucleon spectrum that is
absent in Figure~\ref{fig:spin_840} is a low-lying Roper resonance, a
positive-parity excitation \textit{below} the lowest-lying
negative-parity state.  This feature of a quark-model type ordering of
the low-lying positive- and negative-parity excitations is a feature
of most of the lattice calculations discussed above.  The exception to
this is the calculation of the $\chi$QCD group, using overlap fermions
on an ensemble generated using domain-wall fermions\cite{Liu:2014jua};
their calculation, together with the results of other groups, is shown
in Figure~\ref{fig:roper}. It has been argued that the lattice
calculations of the Roper are incomplete, lacking the multi-hadron
operators as we discuss below, but not inconsistent with the observed
spectrum\cite{Leinweber:2015kyz}.

\subsection{Multi-hadron Spectrum and the nature of resonances}
Lattice QCD is formulated in Euclidean space, and the energies
entering into the spectral decomposition of eqn.~\ref{eq:matrix} are
real. On a finite, discretized Euclidean lattice, the spatial momenta
are quantized, and the calculated discrete spectrum, that is the
energies of eqn~\ref{eq:2pt}, should include two- and higher-body
scattering states.  Indeed, even at the unphysically large quark
masses used in Figure~\ref{fig:spin_840}, many of the higher
excitations are above decay thresholds. For non-interacting particles,
the energies are given by the symmetries of the volume in which we are
working and the allowed spatial momenta.  The finite, periodic spatial
volume forces those hadrons to interact thereby shifting the
energies from their non-interacting values. For the case of elastic
scattering, the so-called L\"{u}scher method enables the shift in
energies at a finite volume to be related to the infinite volume phase
shift\cite{Luscher:1986pf,Luscher:1990ux}, shortly thereafter extended
to states with non-zero total momentum\cite{Rummukainen:1995vs}.  In
the meson sector, the precise calculation of the momentum-dependent
phase shifts for states such as the $\rho$ meson in $I=1 \, \pi\pi$
scattering has now been accomplished\cite{Dudek:2012xn}, and the
formalism has been extended to the extraction of the
momentum-dependent amplitudes for inelastic
scattering\cite{He:2005ey,Guo:2012hv,Hansen:2012tf,Briceno:2012yi}.

Analogous calculations in the baryon sector are less advanced.  For
most of the calculations discussed above, the interpolating operators
were constructed from three quarks, albeit with quite elaborate
orbital structures.  
The two-hadron energy levels that should be seen in the spectrum are in
general absent, and their non-observation is attributed to the volume
suppression of single-hadron operators to multi-hadron states. Several
groups have included multi-hadron or five-quark operators in their
basis\cite{Kiratidis:2015vpa,Verduci:2014csa}.  Notably, one group has
included interpolators of the form ${\cal O}_{N \pi}(\vec{p}) =
N^{i}(\vec{p}) \pi(\vec{0})$, where $N^i, i = 1,2,3$, are the three
local interpolators of eqn.~\ref{eq:Ninterp}, and observe $N\pi$
states as expected, but also the presence of an additional
energy level in the spectrum.  A different approach, applicable to
transitions near threshold\cite{McNeile:2002fh}, has been applied to
the decays of decuplet baryon resonances to an octet baryon-pion final
state\cite{Alexandrou:2015hxa,Alexandrou:2015qdp}.  However, the
application of the full panoply of the L\"{u}scher method is currently
limited.

\section{Electromagnetic transitions in $N^*$ resonances}
The formalism to extract infinite-volume transition matrix elements
between states containing two or more hadrons from Euclidean space
lattice calculations has only recently been
developed\cite{Briceno:2014uqa,Briceno:2015csa,Briceno:2015tza}, and
applied to meson transitions\cite{Briceno:2015dca}; it will be
presented workshop\cite{raul}.  Instead, I will backtrack somewhat and
talk about calculations of electromagnetic properties in which both
the incoming and outgoing particles are treated as single-particle
states.

Hadronic matrix elements in lattice QCD are computed through the
calculation of three-point functions:
\begin{eqnarray}
  C_{\rm 3pt}(t_f, t; \vec{p}, \vec{q})  & = & \sum_{\vec{x}}\sum_{\vec{y}} \langle 0 \mid {\cal
    O}_1(\vec{x}, t_f) J(\vec{y}, t) \bar{\cal O}_2(0) \mid 0 \rangle
  e^{-i \vec{p} \cdot \vec{x}} e^{-i \vec{q}\cdot \vec{y}} \nonumber\\
& \rightarrow & \langle 0 \mid {\cal O}_1 \mid N_1, \vec{p} \rangle \langle N_1, \vec{p}\mid \bar{\cal O}_2 \mid 0 \rangle \langle N_1, \vec{p} \mid J
\mid N_2, \vec{p} + \vec{q} \rangle \nonumber\\
 & & \times e^{- E_1(\vec{p}) (t_f - t)} e^{-
  E_2(\vec{p} + \vec{q}) t},\label{eq:3pt}
\end{eqnarray}
with $Q^2 = (E_2(\vec{p}) - E_2(\vec{p}+\vec{q}))^2 - \vec{q}^2$, and
where we treat ${\cal O}_1$ and ${\cal O}_2$ as ideal interpolating
operators for the incoming and outgoing states, obtained, for example,
from the eigenvectors obtained in the variational
method\cite{Shultz:2015pfa}.

\begin{figure*}
  \centering
  \includegraphics[width=0.45\textwidth]{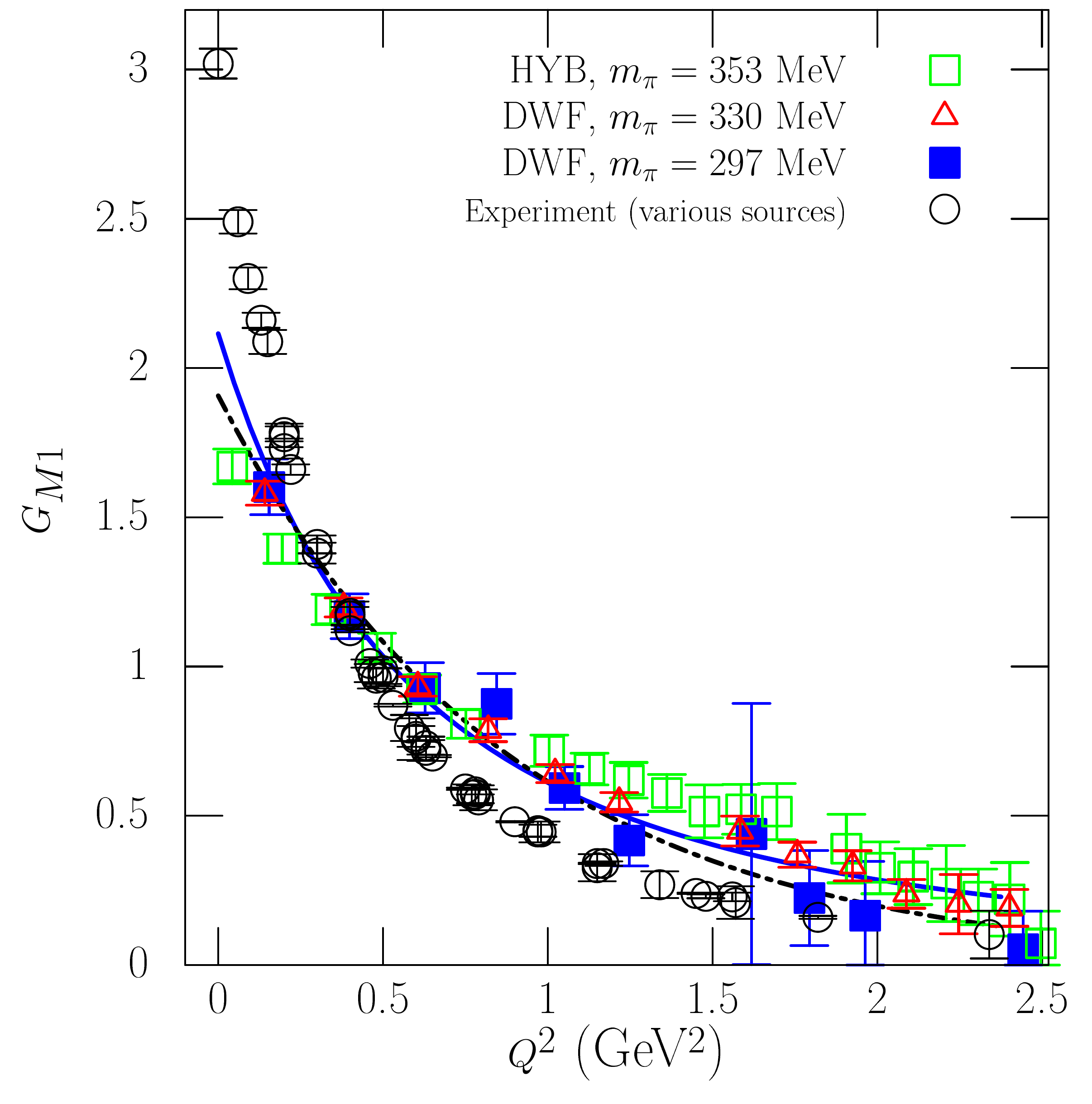}
  \includegraphics[width=0.45\textwidth]{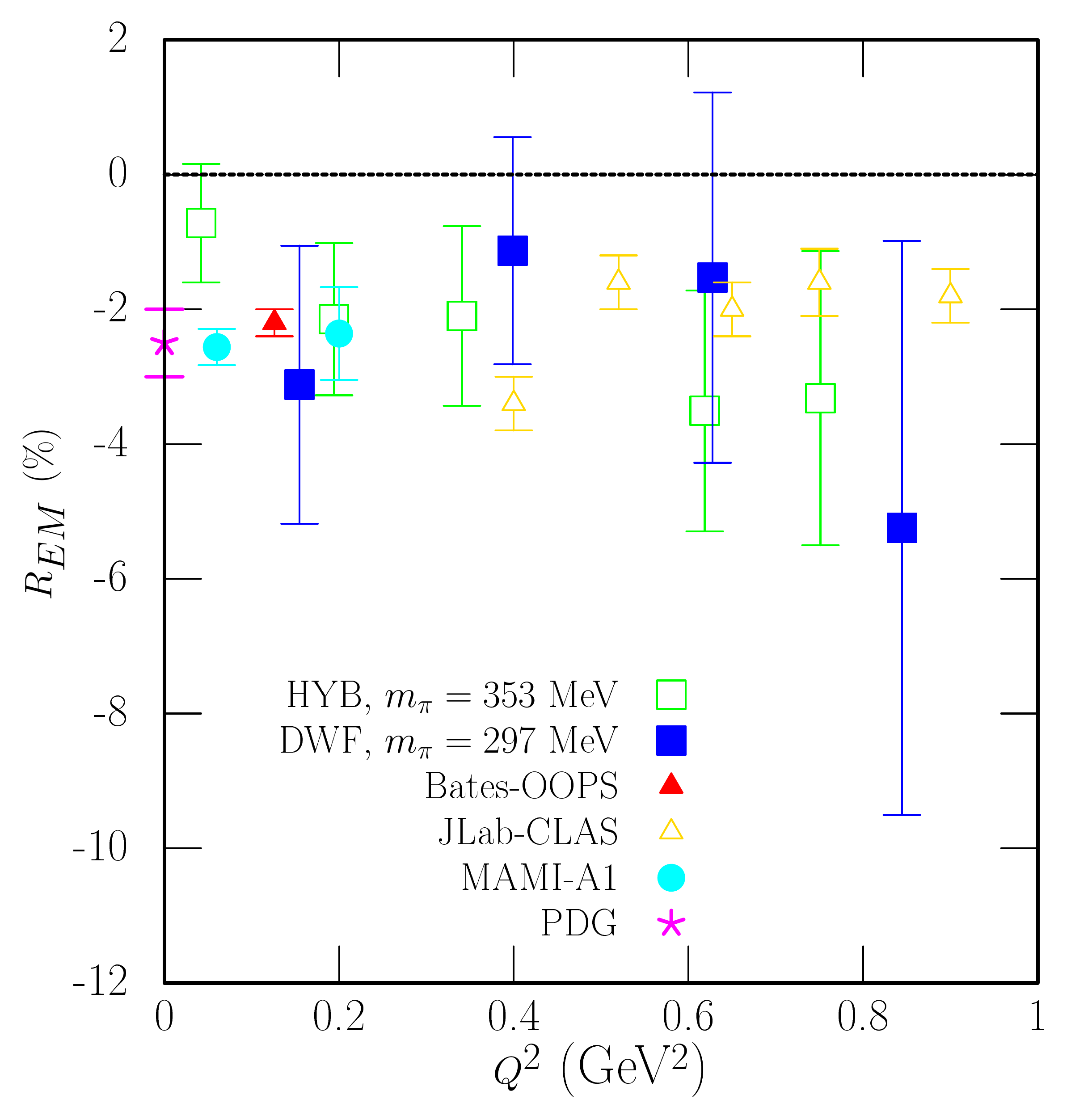}
  \caption{The left-hand plot, taken from
    ref.~\protect\cite{Alexandrou:2010uk} shows the magnetic dipole
    form factor $G_{M1}(Q^2)$ for the $N$ to $\Delta$ transition.  The
    right-hand plot, taken from the same paper, shows the ratio
    $R_{EM} = - G_{E2}(Q^2)/G_{M1}(Q^2)$.}
  \label{fig:delta}
\end{figure*}
The most widely studied transition is that to the $\Delta$
resonance\cite{Alexandrou:2006mc,Alexandrou:2010uk}, though there was
an earlier investigation of the $P_{11}$ Roper
transtion\cite{Lin:2008qv}. The former is of particular interest both
because there has been considerable experimental effort, and because a
non-zero quadrupole moment can provide important information about the
structure nucleon inaccessible from the proton form factor.  A recent
example of such a calculation is shown in Figure~\ref{fig:delta}, for
both the magnetic dipole transition form factor and for the ratio of
the quadrupole to dipole form factors.  As noted earlier, it must be
emphasised that the interpretation of infinite-volume matrix elements
for multi-hadron states requires considerable theoretical
work\cite{Agadjanov:2014kha,Agadjanov:2016aip}, and the application is
computationally demanding.

  As the momenta in eqn~\ref{eq:3pt} increase, lattice calculations
  of the form factors become increasingly demanding, both because of
  decreasing signal-to-noise ratios and because of increasing
  discretisation errors.  There have been several ideas that aim to
  overcome these issues\cite{Lin:2010fv,Bali:2016lva} to enable the
  direct calculation of the $Q^2$ dependence of the form factor, but another
  approach, applicable to hard, exclusive processes, is through the
  calculation of the hadronic wave functions for both the nucleon and
  for its excitations\cite{Bali:2015ykx,Braun:2014wpa}; that approach
  is described elsewhere at this workshop\cite{braun}.

\section{Summary}
There has been far-reaching theoretical and computational progress
aimed at understanding the excited state spectrum of QCD through
lattice calculations.  Much of the computational work has been
centered on the meson sector, where the computational are considerably
reduced.  However, with the increasing computational resources, and
our increasing theoretical understanding, these ideas will be applied
to the $N^*$ spectrum, enabling lattice calculations to truly confront
experiment.

\begin{acknowledgements}
The author would like to thank his colleagues in the \textit{HadSpec
  Collaboration} for discussions and collaboration on some of the
work.  I am grateful to the authors of refs.~\cite{Liu:2014jua} and
\cite{Alexandrou:2010uk} for use of their figures.  This material is
based upon work supported by the U.S. Department of Energy, Office of
Science, Office of Nuclear Physics under contract DE-AC05-06OR23177.
\end{acknowledgements}

\bibliographystyle{spbasic}
\bibliography{richards.bib,richards_raul.bib}   

\end{document}